\documentclass[prb,twocolumn,showpacs,preprintnumbers,amsmath,amssymb]{revtex4}
\usepackage{graphicx}
               
\usepackage{graphicx}

\newcommand{\A}{\underline{A}}
\newcommand{\B}{\underline{B}}
\newcommand{\C}{\underline{C}}

\begin{document}

\title{Magnetism of two-dimensional defects in Pd: \\
       stacking faults, twin boundaries and surfaces}

\author{Simone S. Alexandre}
\author{Eduardo Anglada}
\affiliation{Departamento de F\'{i}sica de la Materia Condensada,
Universidad Aut\'onoma de Madrid, 28049 Madrid, Spain.}

\author{Jos\'e M. Soler}
\author{F\'elix Yndurain}
\email[E-mail: ]{felix.yndurain@uam.es}
\affiliation{Departamento de F\'{i}sica de la Materia Condensada,
Universidad Aut\'onoma de Madrid, 28049 Madrid, Spain,}
\affiliation{Instituto de Ciencia de Materiales Nicol\'as Cabrera,
Universidad Aut\'onoma de Madrid, 28049 Madrid, Spain.}

\date{\today}

\begin{abstract}
   Careful first-principles density functional calculations reveal the 
importance of hexagonal versus cubic stacking of closed packed planes 
of Pd as far as local magnetic properties are concerned. 
   We find that, contrary to the stable face centered cubic phase, 
which is paramagnetic, the hexagonal close-packed  phase
of Pd is ferromagnetic with a magnetic moment of 0.35\ $\mu_{B}$/atom. 
   Our results show that two-dimensional defects with local hcp stacking, 
like twin boundaries and stacking faults, in the otherwise fcc Pd structure, 
increase the magnetic susceptibility. 
   The (111) surface also increases the magnetic susceptibility and it
becomes ferromagnetic in combination with an individual stacking fault 
or twin boundary close to it.
   On the contrary, we find that the (100) surface decreases the tendency 
to ferromagnetism.
   The results are consistent with the magnetic moment recently observed 
in small Pd nanoparticles, with a large surface area and a high 
concentration of two-dimensional stacking defects. 
\end{abstract}



\maketitle

\section{\label{sec:introduction}Introduction}

   Despite their narrow $d$-bands and high densities of states (DOS) 
at the Fermi level, which favor magnetism, there are only three
ferromagnetic transition metals in nature.
   Palladium is paramagnetic in its stable fcc structure, 
but with a very high magnetic susceptibility. 
   Several calculations have shown that its Fermi level lies just
above a large peak in the DOS, at the top of the $d$-bands.
   The DOS at the Fermi-level, of $\sim 1.1$ states per spin, eV and atom
(Chen {\it et al}. \cite{chen} and references therein) is almost
high enough, but not quite so, to fulfill the Stoner criterion for 
itinerant magnetism, since the Stoner exchange parameter is 
$\sim 0.73$ eV \cite{jan,tak}. 
   The calculations have also shown that Pd, in its fcc crystal structure, 
becomes ferromagnetic by increasing the lattice constant by just a few 
percent \cite{mor}. 
   All these results have stressed the subtle balance of magnetism in Pd.
 
   Therefore, we suggest that variations in the atomic arrangement can 
induce changes in the density of states at the Fermi level that, in turn, 
can induce magnetism. 
   In this direction, it has been proposed that monatomic Pd nanowires
are ferromagnetic, either in their energetic \cite{del,alex} or
thermodynamic \cite{tos} equilibrium length.
   Also, recent experimental results \cite{sam} indicate that fcc Pd 
nanoparticles with stacking faults and twin boundaries present 
ferromagnetism. 
   Other experiments \cite{shi} on small Pd nanoparticles also indicate 
the existence of a hysteresis loop which, in this case, is interpreted 
as due to a non zero magnetic moment at the surface atoms. 

   Stacking faults and twin boundaries are very common two-dimensional 
defects in face centered cubic (fcc) metals as well as in diamond and 
wurtzite-structure semiconductors. 
   Their abundance is partly due to their low energy of formation,
since they preserve the local geometry and close packing.
   Their importance in the mechanical properties, like 
hardness and brittleness are recognized since many years. 
   Recently, the role played by the stacking of (111) layers has been 
emphasized in connection with magnetic orientation and magnetic ordering 
of thin layers and superlattices of magnetic metals~\cite{par,zhe,pie}.
   The growth of Co on top of the Cu(111) surface has been extensively 
studied experimentally, and the correlation between stacking pattern and 
magnetic properties has been reasonably well established. 

   At this stage it seems worth studying the effect of the stacking 
sequences and of the surfaces on the electronic and magnetic properties 
of Pd.
   With these ideas in mind, we have calculated total energies and magnetic 
moments of Pd in both the fcc and hexagonal close-packed (hcp) phases,
as well as in surfaces and near two-dimensional stacking defects like 
intrinsic and extrinsic stacking faults, and twin boundaries.
   

\section{\label{sec:methodology}Methodology}

   All our calculations are performed within density functional 
theory \cite{kohn} (DFT), using either the local density approximation 
\cite{lda} (LDA) or the generalized gradient approximation 
\cite{pbe} (GGA) to exchange and correlation.
   Most of the calculations were obtained with the SIESTA~\cite{ord,sol} 
method, which uses a basis of numerical atomic orbitals~\cite{san} 
and separable~\cite{kle} norm conserving pseudopotentials~\cite{tro} 
with partial core corrections~\cite{lou}. 
   To generate the pseudopotentials and basis orbitals, we use a
Pd configuration $4d^9 5s^1$, since we have checked that it leads
to better transferability and bulk properties than the ground state 
configuration $4d^{10} 5s^0$.
   After several tests we have found satisfactory the standard 
double-$\zeta$ basis with polarization orbitals (DZP) which has been used 
throughout this work.  
   The convergence of other precision parameters was carefully checked.
   The range of the atomic basis orbitals was obtained using an energy 
shift~\cite{sol} of 50 meV. 
   The real space integration grid had a cut-off of 500 Ryd, while around 
9000 $k$ points/atom$^{-1}$ were used in the Brillouin zone sampling. 
   To accelerate the selfconsistency convergence, a broadening of the 
energy levels was performed using the method of Methfessel and Paxton
\cite{met} which is very suitable for systems with a large variation of 
the density of states at the vicinity of the Fermi level.

   It is necessary to mention that most of the energy differences
between paramagnetic and ferromagnetic solutions in Pd structures
are extremely small, what requires a very high convergence in all
precision parameters and tolerances, and specially in the number
of $k$ points.
   It must be recognized, however, that the basic DFT uncertainty
is probably larger than those energy differences, so that it is 
not really possible to determine reliably whether a particular
defect or structure is para or ferromagnetic.
   Still, we think that it is possible to find reliably the relative
tendency towards magnetism of different structures.
   In particular, $E(M)$ curves, of total energy versus 
total magnetic moment, provide an excellent tool to study
the tendency to magnetism in different systems: independently of
whether the systems are para or ferromagnetic, one may determine
if a defect or a surface has a smaller or a larger tendency to
magnetism than the bulk, depending on which of the
two $E(M)$ curves is higher.
   In addition, the selfconsistent convergengy is considerably faster
at constant magnetic moment, so that it is possible to determine 
reliably (for a given functional) relative energies as small as a
fraction of an meV.

\section{\label{sec:bulk}Bulk crystals}

   The first mandatory system to be considered is the perfect bulk
crystal in its experimentally-stable fcc phase.
   In principle, the GGA functional goes beyond the LDA and it is 
generally considered to be more accurate and reliable.
   However, in the case of Pd, we find that the GGA gives a lattice 
constant of 3.99 {\AA}, 2.5\% larger than experiment, and a 
ferromagnetic ground state with a magnetic moment of $0.4\mu_{B}/$atom 
and an energy 4.5 meV/atom below the paramagnetic phase (Figure 1). 
\begin{figure}[h]
\includegraphics[width=\columnwidth,clip]{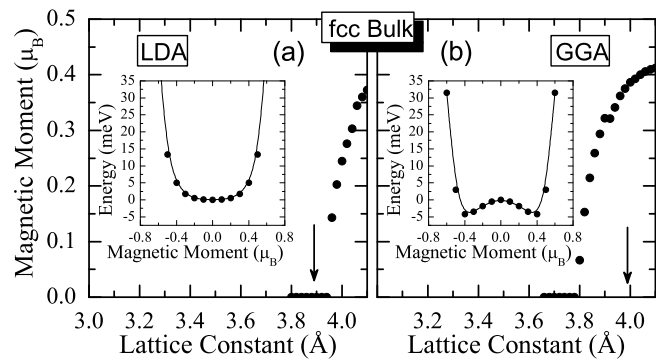}
\caption{ 
   Magnetic moment versus lattice constant for fcc Pd obtained with 
the SIESTA code and the LDA and GGA functionals.
   The insets indicate the variation of the total energy with 
magnetic moment at the corresponding equilibrium lattice constants,
indicated by arrows.}
\label{fig:fccbulk}
\end{figure}
   Since this energy difference is very small, and to discard that the 
ferromagnetic phase is favoured by the pseudopotential or basis set used,
we have reproduced~\cite{alex} this result using two other DFT methods: 
the pseudopotential plane wave code VASP \cite{vasp} and the all-electron
augmented plane wave method WIEN \cite{wien}, both with the same GGA 
functional \cite{pbe}.
   Previous GGA calculations by Singh and Ashkenazi \cite{singh},
using a different functional, also found a lattice constant larger 
than the experimental one, but did not address the subtle
ferromagnetic-paramagnetic transition with sufficient detail.

   On the other hand, within the LDA we find a paramagnetic ground 
state and a lattice constant of 3.89 {\AA}, both in agreement with the 
experimental values and with previous LDA calculations of Moruzzi and 
Marcus \cite{mor}.
   These results suggest that the GGA is not necessarily more reliable
to study magnetism in Pd.
   In fact, it is not feasible to study the possible existence of
magnetic defects in fcc Pd when the bulk result is already ferromagnetic.
   Therefore, we have chosen the spin-dependent LDA to perform most of 
the remaining calculations in this work, although GGA results have been 
obtained also as a check in some cases.

\begin{figure}[h]
\includegraphics[width=1.0\columnwidth,clip]{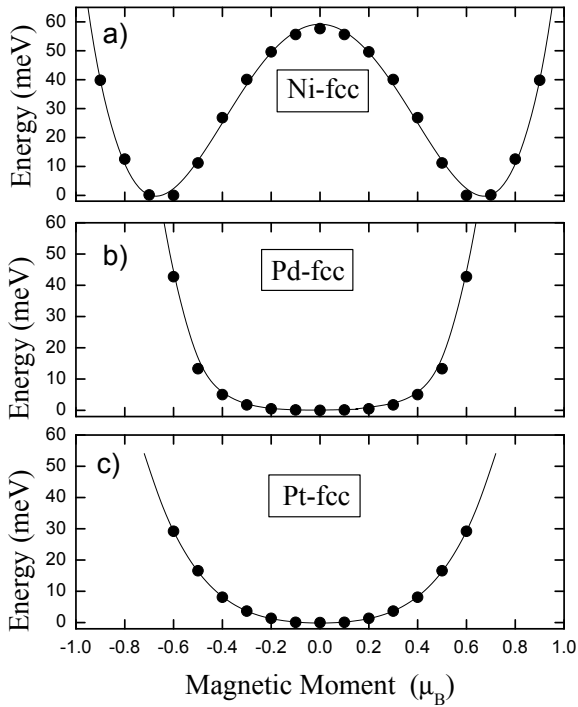}
\caption{
   Total energy versus magnetic moment for three transition 
metals in the same column of the Periodoc Table, calculated in the LDA. 
   Note the flatness of the Pd curve, responsible of the large magnetic 
susceptibility of this metal. }
\label{fig:NiPdPt}
\end{figure}

   The magnetic susceptibility $\chi$ of the magnetic moment $M$ to an
external magnetic field $H$ is
\[ \chi= \left( \frac{\partial M}{\partial H} \right)_{H=0}
  = \left( \frac{\partial^{2}E}{\partial M^{2}} \right)_{M=0}^{-1} \]
   Thus, the flatness of the $E(M)$ curve in Figure \ref{fig:fccbulk} 
implies a very large susceptibility, even within the LDA.
   To compare with the other transition metals in the same column of the 
Periodic Table we have calculated the variation of energy with magnetic 
moment for Ni and Pt. 
   The results are shown in Figure \ref{fig:NiPdPt}. 
   We immediately observe a clear ferromagnetic and paramagnetic behavior 
of Ni and Pt respectively, while Pd, as indicated above, is paramagnetic 
but very close to the ferromagnetic transition.

\begin{figure}[h]
\includegraphics[width=1.0\columnwidth,bb=14 40 660 540,clip]{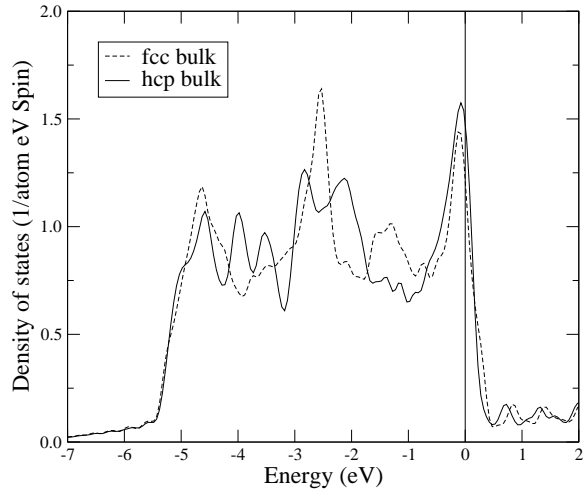}
\caption{
   Calculated electronic densities of states of Pd in the paramagnetic 
phase (i.e. forcing equal spin up and spin down populations) in the fcc 
and hcp crystal structures. 
   The vertical line indicates the Fermi level.
}
\label{fig:densidade}
\end{figure}

   In order to study how magnetism depends on the local geometry and 
stacking of atoms, we have first considered the hcp 
structure as compared to the fcc one. 
   It is known  that the breaking of the cubic symmetry in stacking faults, 
while keeping the number of nearest-neighbor atoms and their bond lengths 
and angles, induces a rearrangement of the $d$ energy bands that can even 
give rise to localized electronic states~\cite{ynd}. 
   The DOS of hcp and fcc structures, in the paramagnetic 
phase, shown in Figure \ref{fig:densidade}, were calculated for the 
same nearest neighbor distance of 2.76 {\AA} and the ideal ratio 
$c/a=(8/3)^{1/2}$ for the hcp structure, since we obtain that 
both structures are stable at this distance, with the hcp $c/a$ ratio 
only 0.75\% larger than the ideal value.
   Several points are worth mentioning: 
   {\it(i)} The $d$ bandwidth is almost identical in both structures
because they have identical nearest-neighbor configurations. 
   {\it(ii)} The shapes of the DOS are very different, 
as a consequence of the breaking of the cubic symmetry: 
in fcc the second-nearest-neighbor atoms are staggered whereas in hcp 
they are eclipsed and the lack of cubic symmetry inhibits the 
$t_{2g}-e_g$ splitting of the $d$ bands. 
   {\it(iii)} The DOS at the Fermi level is larger in 
hcp than in fcc (1.43 versus 1.15 states per atom per eV and per spin).
   This implies that Pd in the hcp structure satisfies the Stoner 
condition  for ferromagnetism: with a Stoner exchange parameter 
$I$=0.73 eV \cite{jan,tak}, we get $I\times D(E_F)=0.84$ for fcc 
and $I\times D(E_F)=1.05$ for hcp.

\begin{figure}[h]
\includegraphics[width=\columnwidth,clip]{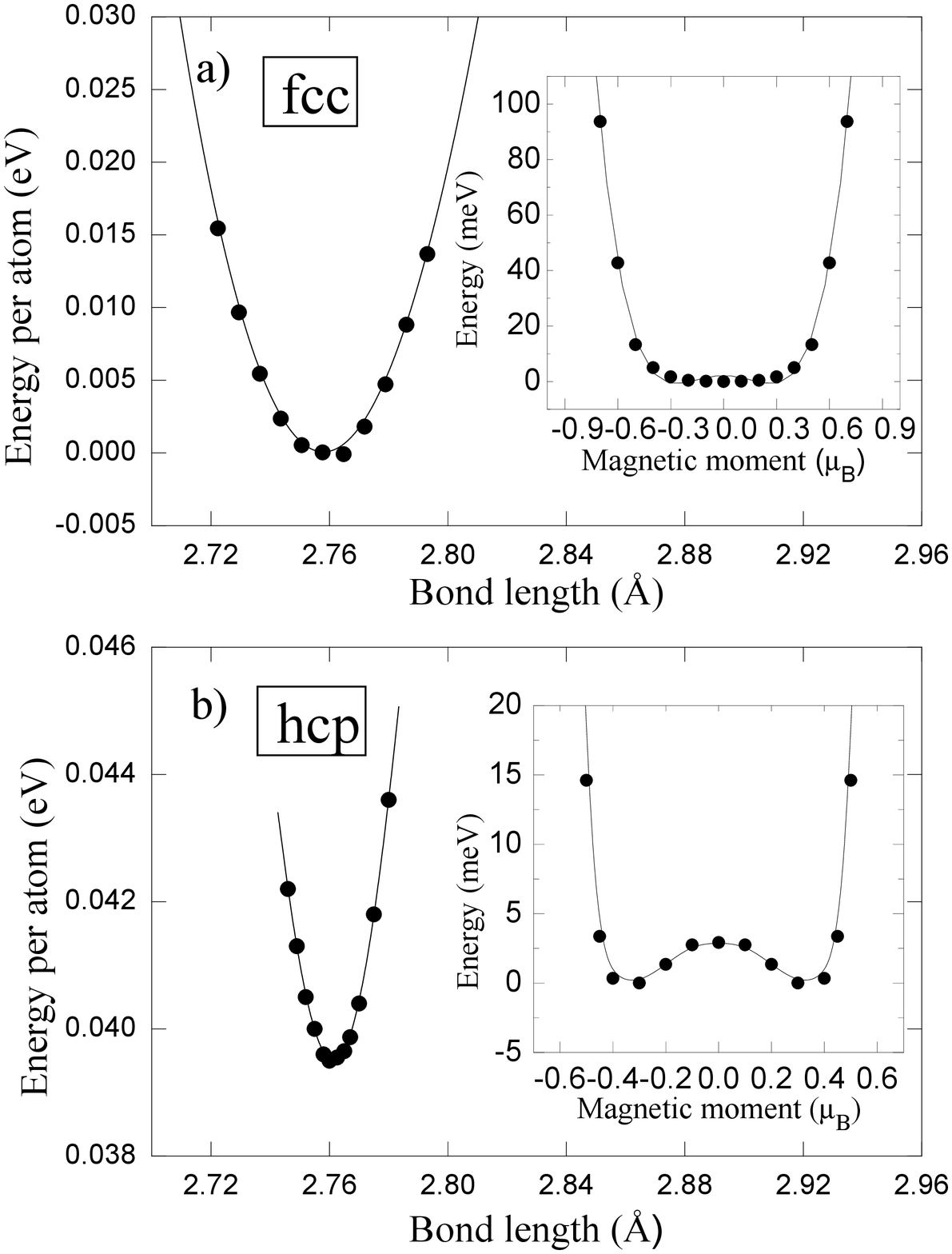}
\caption{
   Calculated total energy for (a) fcc and (b) hcp Pd 
as a function of the nearest neighbor distance. 
   The arrows indicate the minimum (equilibrium) nearest-neighbor distance. 
   The insets at (a) and (b) show the total energy versus 
magnetic moment for the fcc and hcp structures respectively. 
   In these cases the energy origin is at the corresponding minima.
}
\label{fig:hcpfccmaio}
\end{figure}

   We have then calculated the total energy and the magnetic order in 
the hcp phase in the LDA. 
   The results are shown in Figure \ref{fig:hcpfccmaio}. 
   We immediately observe that, contrary to what happens in the fcc 
structure, there is a non zero magnetic moment  of 0.35\ $\mu_{B}$/atom
at the equilibrium interatomic separation for the hcp structure,
$d$=2.76 {\AA}. 
   Like in the fcc phase, the system is close to a magnetic-nonmagnetic 
transition but, in this case, in the ferromagnetic side, with the
ferromagnetic hcp phase approximately 1.7 meV lower in energy than the 
paramagnetic one.
   The transition from para to ferromagnetism, as a function of the 
lattice constant, is abrupt, like in the fcc case, although we cannot 
assess with enough confidence whether it is first or second order 
\cite{mor,mor2}. 

   Finally, the hcp structure is 4.0 meV higher in energy than the fcc.
   This ordering is in agreement with experiment, and we reproduce it
also with the GGA, but it is in contradiction 
with the calculations of Huger and Osuch \cite{hcp} that reported
an hcp structure lower in energy, but also ferromagnetic.

\section{\label{sec:defects}Bulk defects}

   Hexagonal (111) closed packed planes of atoms can be stacked in 
different ways, giving rise to different structures. 
   If we label the three possible positions of the atoms as A, B and C,
the fcc stacking is ...ABCABC... and that of hcp is 
...\underline{ABABAB}...\cite{kittel}, where the layers with hexagonal
symmetry are undelined.
   Different defects can be generated in the fcc structure:
   The intrinsic and extrinsic stacking faults have two hexagonal layers,
with stacking sequences ...ABC{\A\C}ABC... and ...ABC{\A}C{\B}CABC...
respectively.
   The twin boundary has a single hexagonal layer with ...ABC{\A}CBA... 
stacking.

   Previous model calculations \cite{ynd} at stacking faults
of transition metals indicate an important perturbation of the local 
densities of states  from that of the perfect fcc lattice. 
   The presence of a hexagonal stacking of layers induces localized 
electronic states \cite{ynd,vaz} and an enhancement of the DOS 
at the top of the valence band. 
   These calculations suggest possible variations of the magnetic 
properties around the extended defect.  
   We have performed calculations of several packing sequences, 
in supercells containing various numbers of fcc layers, to study to what 
extent different local configurations can give rise to magnetic moments. 
   The results of the calculations, both in the perfect crystals and 
in the defects, are independent (within less than 1\%) of the initial 
input moment. 

\begin{figure}[h]
\includegraphics[width=1.0\columnwidth,clip]{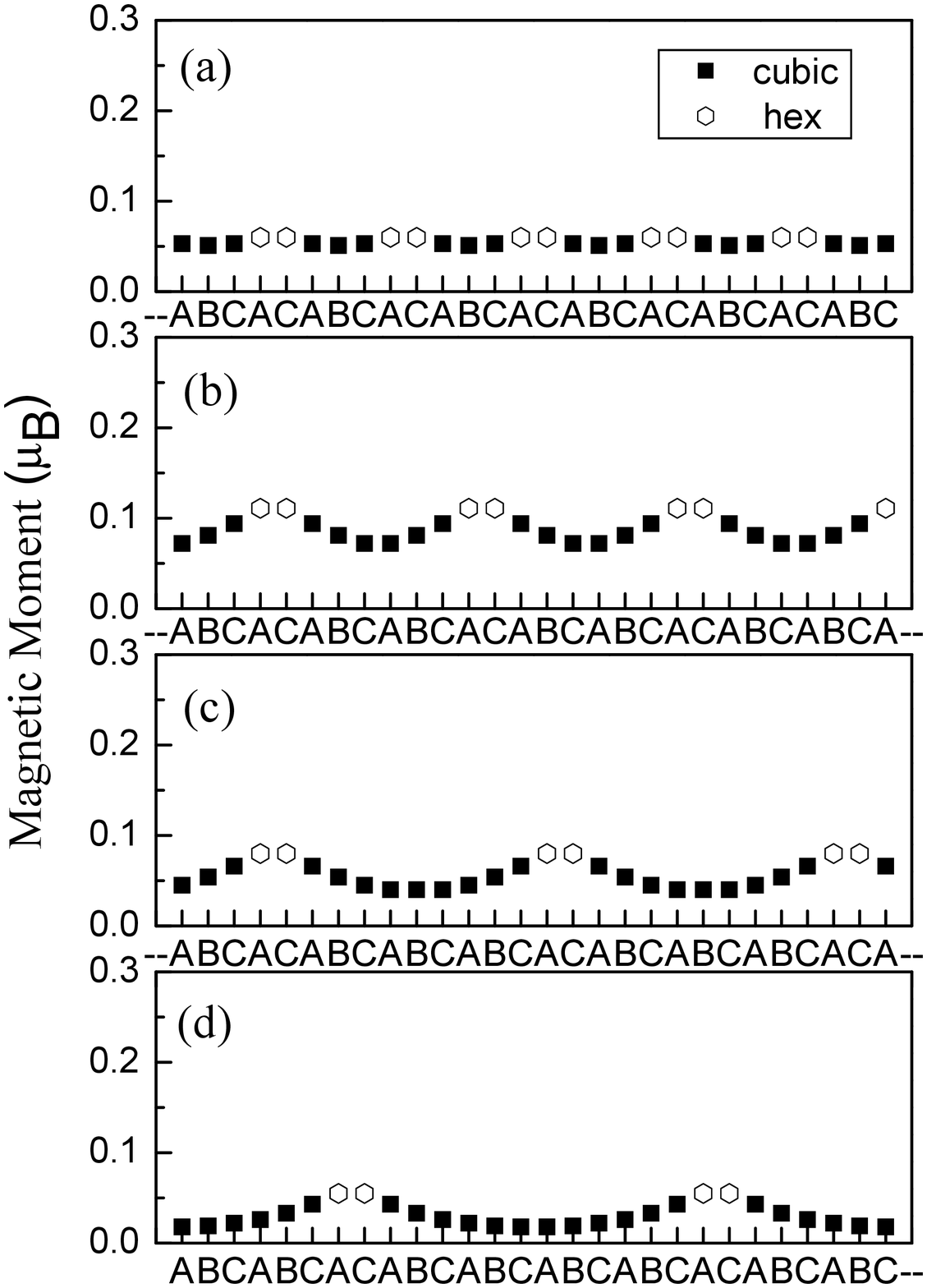}
\caption{
   Local magnetic moment in supercells of different thickness, 
containing a single intrinsic stacking fault between fcc layers,
calculated with the LDA.}
\label{fig:superceldas}
\end{figure}

   Most of the calculated supercells have a finite magnetic moment
and some are shown in Figure~\ref{fig:superceldas}. 
   However, its variation with the supercell thickness is rather 
complex and nonmonotonic. 
   We suspect that this complex behaviour is due to an oscillatory 
component of the magnetic coupling between the neighbouring stacking 
faults, like that observed in superstructures of magnetic slabs 
sandwiched between nonmagnetic metals~\cite{super}.
   Although we have not been able to stabilize any antiferromagnetic 
solution in double-size supercells, this may be due to the size
limitations and to the difficulties of convergence.
   As expected, the magnetic moments are smaller in fcc than in hcp 
layers, but they are nevertheless far from negligible. 
   The interplay between the magnetism of the hexagonal layers and 
the paramagnetism of the cubic ones, both close to a paramagnetic to 
ferromagnetic transition, is very subtle. 
   What is important is that the hexagonal layers have a
tendency to become magnetic and they induce magnetic moments at the 
atoms in the cubic layers. 
   This is due first, to the large magnetic susceptibility in fcc Pd, 
where the hexagonal layers act as magnetic impurities.
   And second, because of the two-dimensional character 
of the defects, any magnetic perturbation decays very slowly with 
distance, like in a pseudo one-dimensional metal. 
   In other words, there is a long-range RKKY-like interaction between 
the hcp layers through the intermediate fcc ones. 

\begin{figure}[h]
\includegraphics[width=\columnwidth,clip]{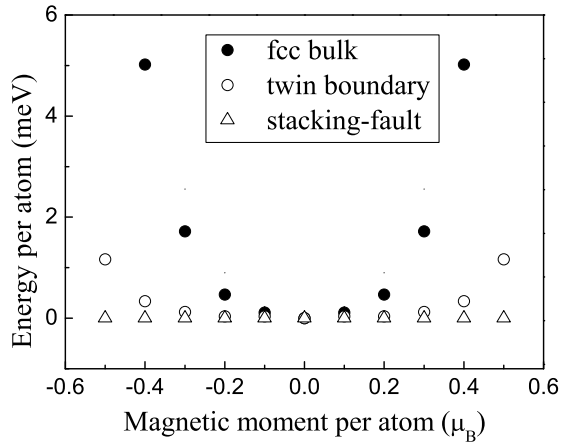}
\caption{
   Total energy as a function of the magnetic moment 
for differents two dimensional defects of Pd,
with local hexagonal symmetry, compared to the bulk fcc lattice. 
   The energy at $M=0$ has been shifted to a common value. 
The defects are separated by 6 fcc layers like in Figure 5(b). 
   The stacking fault curve has a minimum of -0.2 meV
at $M \simeq 0.09 \mu_B$/atom, 
in agreement with Figure \ref{fig:superceldas}b,
which is not noticeable at the figure scale.
}
\label{fig:defectos}
\end{figure}

   Given the size limitations of our calculated supercells, and the
nonmonotonic magnetic moment with increasing supercell thickness, it is not
possible to conclude confidently whether isolated planar defects in fcc Pd
have a finite magnetic moment, within the LDA.
   However, Fig.~\ref{fig:defectos} clearly shows that 
all the hexagonal defects have a larger tendency to magnetism than the
bulk fcc lattice, as indicated by their respective $E(M)$ curves.
   This is hardly surprising, given the ferromagnetic character of hcp Pd.
   On the other hand, all the supercells are of course strongly 
magnetic within the GGA, with a larger tendency to magnetism
(lower $E(M)$ curve) than the bulk fcc crystal.
   Therefore, it is perfectly possible that an isolated stacking
fault in fcc Pd is indeed magnetic in nature.

\section{\label{sec:surfaces}Surfaces}
   
\begin{figure}[h]
\includegraphics[width=1.0\columnwidth,clip]{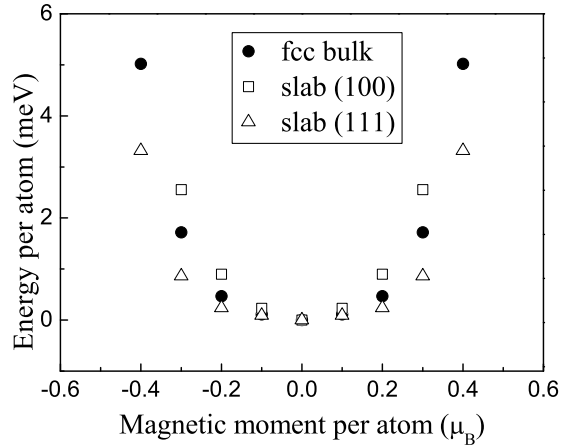}
\caption{
   Total energy within the LDA versus magnetic moment of bulk fcc Pd 
and of 9-layer slabs with surfaces oriented in the (100) and (111) 
directions. 
   The energy at $M=0$ has been shifted to a common value.
}
\label{fig:slabEnergy}
\end{figure}

\begin{figure}[h]
\includegraphics[width=0.45\textwidth]{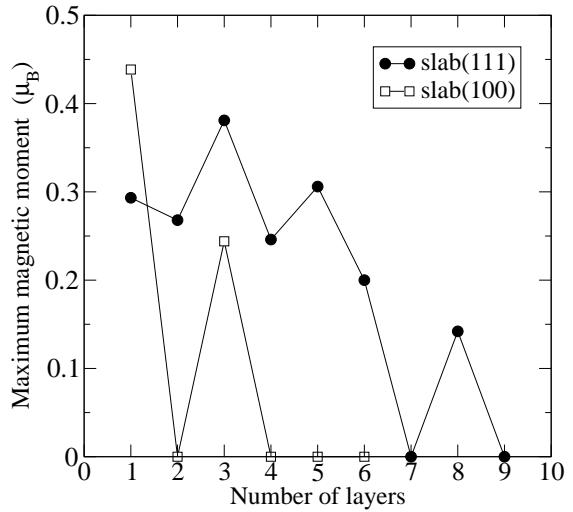}
\caption{
   Largest local magnetic moment, as a function of the 
number of layers, in slabs with surfaces oriented in the (100) and 
(111) directions.}
\label{fig:slabMoment}
\end{figure}

   Recent experimental results \cite{shi} on Pd nanoparticles have
been interpreted as magnetism at (100) surfaces. 
   In principle, surface magnetism is plausible in general because the
lower coordination of surface atoms favors narrower bands and
larger densities of states.
   In practice, however, surface relaxation and reconstruction may
contract the surface bond lengths and more than compensate for the
lower coordination.
   We have then calculated the electronic structure of finite Pd slabs
in the (100) and (111) orientations, after carefully relaxing their 
geommetry. 
   The results of the calculated total energy versus magnetic moment 
are shown in Figure~\ref{fig:slabEnergy} for the largest calculated 
thickness. 
   This general tendency is consistently obtained for sufficiently
thick slabs, but the thikness dependence of the total magnetic moment,
shown in Fig.~\ref{fig:slabMoment} is not monotonic, as in the bulk
supercells.

   We observe in Fig.~\ref{fig:slabEnergy} that the minimun energy is 
at zero magnetic moment and therefore the two surfaces are paramagnetic. 
   Moreover, the $E(M)$ curve of the (100) surface is higher than that 
of the bulk, showing that this surface is less prone to magnetism than 
the bulk.
   On the other hand, the (111) surface, although also paramagnetic
within the LDA, has a larger susceptibility than the bulk, what
makes plausible that it may be magnetic in nature.

\begin{figure}[ht]
\includegraphics[width=1.0\columnwidth,clip]{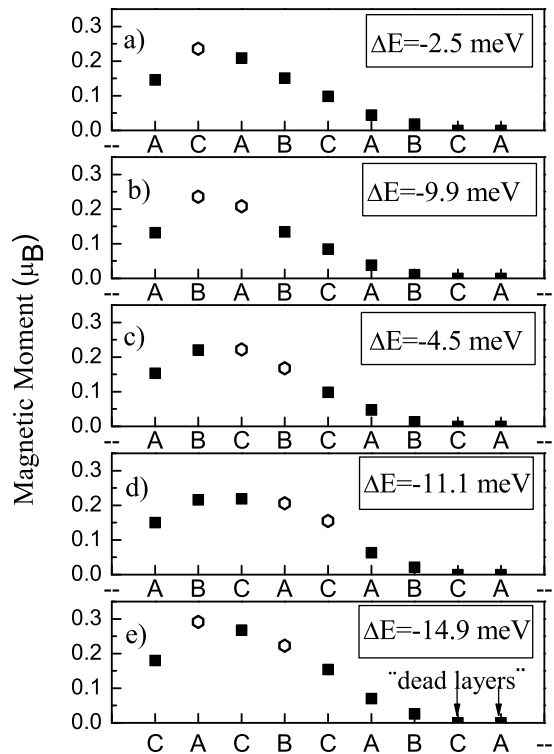}
\caption{
   Local magnetic moments for diferents stackings of hexagonal and
cubic planes in a slab of 9 layers. 
   (a) Twin boundary. 
   (b), (c), and (d) Intrinsic stacking faults, with the hcp layers 
in differents positions. 
   (e) Extrinsic staking fault. 
   Squares and hexagons represent cubic and hexagonal layers, respectively. 
   The distance between the two rightmost layers was fixed to a small
value, in order to kill their tendency to magnetism and thus to
simulate the bulk.
   The energies reported are the difference between the total energy of 
the slab, minus those of the defect in the bulk and of the unfaulted slab.
   The formation energies in the bulk are 33.9, 74.7 and 73.0 meV for the
twin boundary, intrinsic stacking fault, and extrinsic stacking fault, 
respectively.
}
\label{fig:isolStak}
\end{figure}

\begin{figure}[ht]
\includegraphics[width=0.4\textwidth]{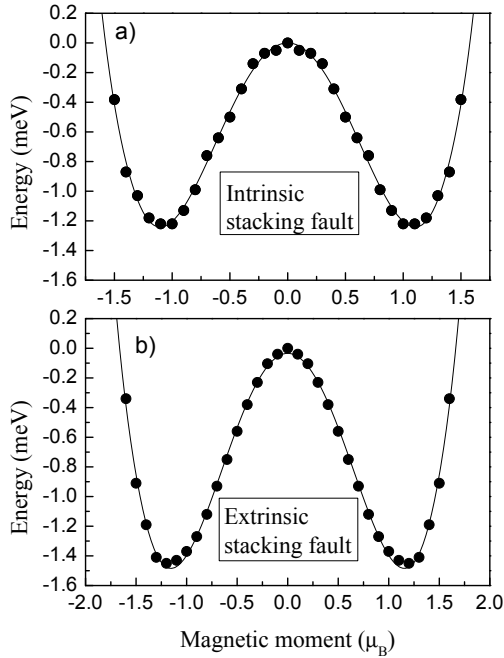}
\caption{
   Calculated total energy, as a function of the magnetic 
moment, for two different two-dimentional defects
close to a (111) surface (shown in Figure \ref{fig:isolStak} (c) and (e)).
}
\label{fig:Slab9-ExM}
\end{figure}

   We have found that bulk planar defects, as well as (111) surfaces, 
have a larger tendency to magnetism than the bulk Pd crystal, which is
itself on the verge of ferromagnetism.
   Since both surfaces and defects are in high concentrations in
nanoparticles, and both have in fact been proposed independently
~\cite{sam,shi} as responsible of the observed magnetic moment of 
these particles, it makes sense to consider their combined effect.
   To this end, we have calculated the geometry, energy, and
magnetism, of planar defects close to a (111) surface.
   We use a slab in which the opposite surface is ``magnetically 
passivated'' by imposing a short distance between the first two 
atomic planes, what inmediately kills their local magnetic monent.
   In this way, we ensure that the possible magnetic moment of
the slab is due to the combination of the stacking fault and
a single surface (in fact, this structure penalizes and sets a
lower limit for the appearence of magnetism).
   Still, as shown in Figure~\ref{fig:isolStak}, we find a clear
magnetic moment in all the cases considered.
   This tendency is further demonstrated by their $E(M)$ curves,
presented in Figure \ref{fig:Slab9-ExM} for two cases, which show
unambiguously their magnetic character.
   Furthermore, the total energy of all the slabs is lower than that 
of the defects in the bulk, plus that of the unfaulted slab, implying 
that the surface atracts the defects and that they should therefore
be expected to appear together in nanoparticles, due to
the high concentration of stacking faults and to the small 
space between surfaces. 
   Notice that the extrinsic stacking fault is more stable than the 
intrinsic one both at the surface and at the bulk.

\section{\label{sec:conclusions}Conclusions}

   Our main conclusions can be summarized as follows: 
   {\it(i)} The GGA functional gives an incorrect ferromagnetic
ground state for the fcc Pd crystal. 
   Accordingly, all the defects studied are also magnetic within
the GGA but, obviously, this does not imply that they are magnetic
in nature.
   On the contrary, the simpler LDA gives the correct lattice 
constant and paramagnetic state.
   {\it(ii)} The hcp phase is ferromagnetic, within both the LDA and GGA.
   In the LDA, it has an energy 1.7 meV lower than the
hcp paramagnetic state and 4.0 meV above the fcc phase. 
   {\it(iii)} We cannot determine whether an isolated stacking fault is
magnetic in the LDA, but it certainly has a larger magnetic
susceptibility than the perfect crystal, and might be magnetic
in nature, given the uncertainty between the different functionals.
   {\it(iv)} The free (100) surface is paramagnetic, with a lower 
susceptibility than the bulk crystal.
   {\it(v)} The (111) surface is paramagnetic in the LDA, but it
has a larger susceptibility than the bulk crystal, and it might
be magnetic in nature.
   {\it(vi)} Hexagonal planar defects are atracted towards a (111) 
surface, and they become clearly magnetic when close enough.

   Therefore, (100) surfaces are not good candidates as the origin 
of magnetism in Pd nanoparticles, as they had been proposed~\cite{shi}.
   In contrast, planar stacking defects, (111) surfaces, and specially 
a combination of both, are plausible candidates to present permanent 
magnetic moments and to be responsible for the magnetism observed in Pd
nanoparticles.
   Thus, our results are consistent with experiments in small Pd 
clusters of average diameter 2.4 nm, which are reported 
to display spontaneous magnetization \cite{sam}. 
   High-resolution transmission electron microscopy has shown that a 
high percentage of the particles exhibit single and multiple 
twinning boundaries.
   In addition, the smallness of the spontaneous magnetization seems 
to indicate that only a small fraction of atoms hold a permanent 
magnetic moment and contribute to ferromagnetism.
   Other experimental results \cite{shi} on small Pd particles have 
also shown their ferromagnetic character. 
   Besides, ferromagnetism can also take place in other non ideal 
structures like nanowires \cite{del}. 

   Magnetic anomalies observed experimentally in different Ni \cite{rob} 
and Co \cite{pie} stacking can be interpreted along the lines described 
in this work. 
   The fact that stacking-faults in Pd display non negligible magnetic 
moments, not present in bulk fcc crystal, opens new lines of research. 
   The appearance of magnetism in nominal fcc samples 
should be revisited in view of our results, since so far the possibility 
of magnetism around stacking faults has been overlooked. 
   Also, layer growth of Pd on top of non magnetic substrates, including 
Pd itself, may produce an interesting magnetic phenomenology in connection 
with the stacking structure which, in turn, depends on the method 
used for growth. 
   The study of the dependence of the magnetic properties of Pd grown 
Ag(111) and Pd/Ag multilayers on the stacking sequences is under way and 
will be reported elsewhere.

\begin{acknowledgements}
   We would like to thank A. Hernando, who brought our attention to 
this problem, to \'O. Paz for his assistance in the technical part 
of this work, and to M. Mattesini for his help in the plane wave basis 
calculations. 
   Work supported Spain's Ministery of Science through grants 
BFM2002-10510-E and BFM2003-03372.
\end{acknowledgements}

\end{document}